\documentstyle[11pt,aaspp4]{article}

\begin{document}

\title{The Enrichment History of Hot Gas in Poor Galaxy Groups}

\author{David S. Davis}
\affil{
Center for Space Research NE80-6003, Massachusetts Institute
of Technology, 77 Massachusetts Ave, 
Cambridge, MA 02139-4307}

\author{John S. Mulchaey}
\affil{The Observatories of the Carnegie Institution of Washington,
813 Santa Barbara St.,
Pasadena CA 91101-1292}

\and

\author{Richard F. Mushotzky}
\affil{Laboratory for High Energy Astrophysics,
NASA/GSFC,
Code 662,
Greenbelt, MD 20771}

\slugcomment{Astrophysical Journal accepted}

\begin{abstract}
We have analyzed the ASCA SIS and GIS data for seventeen
groups and determined the average temperature and abundance of the hot
x-ray emitting gas.
For groups with gas temperatures less than 1.5 keV we find
that the abundance is correlated with the gas temperature and luminosity.
We have also determined the abundance of the $\alpha$-elements and
iron independently for those groups with sufficient counts. 
We find that for the cool groups (i.e. kT $<$1.5 keV) the ratio of
$\alpha$-elements to iron is $\sim$1, about half that
seen in clusters. Spectral fits with the S, Si and Fe abundances 
allowed to vary separately suggest the S/Fe ratio is similar
to that seen in clusters while the Si/Fe ratio in groups is half the 
value determined for richer systems. 
The mass of metals per unit blue luminosity
drops rapidly in groups as the temperature drops. 
There are two possible explanations for this decrease. One is that the
star formation in groups is very different from that in rich
clusters. The other explanation is that groups lose much of their
enriched material via winds during the early evolution of ellipticals.
If the latter is true, we find that poor groups will have contributed
significantly (roughly 1/3 of the metals) to the enrichment of the 
intergalactic medium.

\end{abstract}

\twocolumn
\singlespace
\section{Introduction}
Poor groups of galaxies are the most common aggregation of galaxies
known and thus are the likely repository of much of the material in the low
redshift universe (cf Fukugita, Hogan and Peebles\markcite{FHP98}
1998). In a hierarchical
clustering  scenario for the growth of structure in the universe they are
the first steps in the growth of large scale structure. 
These systems provide a unique opportunity to study the processes
which may occur in more massive systems, such as rich clusters, but with
much less complexity.  X-ray data for poor groups allow the investigation
of the enrichment and heating of the intra-group medium, mass profiles
of poor systems, and the global properties of low mass systems. Because these
groups are the most common structure in the universe their properties
should reflect the processes that dominate the production and ejection of
metals for most stellar systems. In this study we focus on the enrichment
history of the group gas.

Early hints that some groups might contain diffuse gas were provided 
by the Einstein Observatory (Biermann, Kronberg \& Madore \markcite{bkm82}1982;
Bahcall, Harris \& Rood \markcite{bhr84}1984); but, the Einstein IPC
could not clearly distinguish
the x-ray emission from the galaxies and the more diffuse emission
of the group. The $ROSAT$ PSPC is better suited to the search for cool
diffuse gas associated with groups. With superior spatial resolution,
sensitivity to less energetic x-rays, wider field of view and lower
internal background than the Einstein IPC, $ROSAT$ has
provided a large sample of groups with diffuse x-ray gas (Henry et
al. \markcite{H95}1995, Pildis, Bregman, \& Evrard \markcite{PBE95}
1995, Mulchaey et al. \markcite{MDM96}1996, Ponman et
al. \markcite{PBEB96}1996).  These data demonstrate that extended
x-ray emission from groups is a rather common property of these
systems. The implied dark matter distribution often extends on scales
much larger than that inferred from the distribution of the bright
galaxies in these groups. In many groups, the x-ray emitting gas is a
major fraction of the total baryonic mass.

While the $ROSAT$ PSPC is well suited for detecting the low surface
brightness diffuse group gas, its most obvious shortcoming is the lack
of spectral resolution. However, the PSPC is quite sensitive around $\sim$ 1
keV, just where the iron L complex can dominate the x-ray emission for
cool systems and, in principle, should be able to measure the iron
abundance in these systems.  Some questions have been raised about the
accuracy of the PSPC abundance determinations (Bauer \& Bregman
\markcite{bb97} 1997). Because the PSPC does lump all the emission
from the iron L region into one spectral feature these concerns may
be warranted.  With its superior  spectral resolution, wider energy
band, and modest spatial resolution, the x-ray telescope ASCA can be
used to better constrain the temperature and abundance of the hot
extended gas. In particular, the ASCA SIS data can resolve lines
and line complexes from various elements in the spectra from the diffuse
gas. Direct comparison of the ASCA and PSPC abundance measurements have the
potential for verifying the accuracy of the much larger PSPC data base.

\section{Observations and Spectral Fits}
The groups analyzed here are from our accepted ASCA program and from 
the public data archive and therefore, do not form a complete sample. 
Table 1 shows the ASCA group data presented in this paper. The table lists 
the group name in column 1, followed by the Right Ascension and
Declination in the next two columns. 
Columns 4 and 5 list the heliocentric radial velocity and the velocity
dispersion of the group. The final column of the table lists the total
blue luminosity of the galaxies in the group, with the total blue magnitude
taken from the RC3 when available. Otherwise, the magnitude is taken
from the original source (e.g. Hickson et al. \markcite{HKA89}1989;
Zabludoff \& Mulchaey \markcite{ZM98}1998).

Prior to extracting the group spectra
the ASCA data are screened using the default criteria to eliminate
periods of high background, SAA passage and earth block. Background spectra
for the SIS and GIS are from the files generated from blank sky data and
are subtracted from the source spectra. In a few cases, the group emission did
not fill the GIS FOV. For these groups we used a local background for
determining the source spectra. The spectra for the global fits were
extracted from the 
center of the group to the edge of the group emission as observed by the
$ROSAT$ PSPC data. When no PSPC data were available for the groups we 
estimate the extent from the smoothed GIS data.

In the cases where we detected x-ray emission from other sources in the field
we excluded a circular region 3\arcmin$\,$ in radius around the source. This
does not remove all the contaminating flux from the extracted
spectra. However, given the observed count rates of the contaminating
point sources in these fields, we expect only a handful of
``contaminating'' counts in the resulting group spectrum. We did not 
exclude the central elliptical galaxy from the group gas spectrum.

The extracted spectra are grouped so that they have a minimum of 25 counts
per channel. The grouped spectra are fit using XSPEC 9.0 with a Raymond-Smith
plasma model with an absorbing column. We fixed the N$_{\rm H}$ to the Galactic
value for one set of fits and repeated the fit with N$_{\rm H}$ free to vary.
In general, this did not affect the fitted parameters significantly except to
increase the error estimates in the variable N$_{\rm H}$ case. In addition to
the spectral fits with the abundance of the individual elements in the
model varying together, for some spectra we also allowed groups of elements
to vary independently using the VRAYMOND model in XSPEC.

We fit the SIS and GIS data jointly to determine the temperature and
abundance of the group gas. The results are listed in table 2 and are
emission weighted values for the group. Temperature or metalicity 
gradients in the groups will complicate the discussion.
The errors given in table 2 are the 90\% confidence errors. 
The Galactic
N$_{\rm H}$ values are from Dickey \& Lockman \markcite{DL90}(1990).
To assure that the x-ray properties of the groups listed in table 2 are uniform
we have re-analyzed the data for several groups that also appear in the
literature. For example, the $ASCA$ data for NGC~5044 and HCG~51 have
been analyzed by Fukazawa et al. \markcite{FMM96}(1996). Our
temperatures and abundances agree with their determinations within the
90\% confidence errors. 
In addition to those we have HCG~62, MKW4s, and MKW9 in common
with the groups analyzed by Hwang \markcite{H97}(1997) and again our derived
temperature and abundance agrees very well with her results.
So, unlike the case with the $ROSAT$ PSPC analysis for groups which showed
wide variations of x-ray properties depending on the analysis method, the
ASCA results seem to be fairly robust. We will compare the ASCA and $ROSAT$
PSPC results in more detail in a later paper (Mulchaey et al.
\markcite{MDM98}1998), but in general we find good agreement between
the ASCA and $ROSAT$ results.

\section{Diffuse Gas in Groups}

The gas temperature in equilibrium systems reflects the depth of the 
potential well. The majority of groups in our sample show a fairly 
narrow range of temperature. The group temperatures peak at around 1
keV which is consistent with the narrow range of masses for poor
groups found by Mulchaey et al.\markcite{MDM96} (1996).
The x-ray temperature of the NGC~3258, NGC~4104 and
MKW~9 groups are about twice the characteristic temperature of the
cooler groups which indicates that these are more massive systems. 
The seventeen groups span a much larger range in metal 
abundance (from $<$0.05 solar to 0.55 solar with a median value of
0.25 solar).  The three groups that we identify as ``hot'' do not
distinguish themselves in metal abundance with all three 
having abundances close to the median value. 

The metal abundance in these groups appears to show a strong
variation with temperature (figure 1), in the sense of increasing metal
abundance with increasing temperature for temperatures below about 1.5
keV. This is very different from the trend seen in richer systems. 
The metal abundance for clusters of galaxies shows either no variation
(Fukazawa et al. \markcite{FMM96} 1996) or  a slow rise in the metal
abundance between 10 -- 2 keV (Arnaud et al \markcite{arbvv92}1992;
Hatsukade \markcite{H89}1989).
With the addition of groups it is clear that below a temperature
of $\sim$1.5 keV the measured Fe abundance drops quickly to about half the
value seen in the hotter groups and continues to drop for the lower
temperature systems. 
A correlation between the temperature and
abundance seems to be present in the data (see figure 1). Using
Kendall's $\tau$ to test for the presence of a 
correlation, we find less than a 3\% chance that a correlation is not 
present in the data. Fitting the data with a power law, we find that
the best fit power law is 
Abundance $\propto$ kT(keV)$^{+2.5\pm 0.94}$, which is quite
different from that found for clusters (Abundance $\propto$
kT(keV)$^{-0.37}$; Arnaud et al. \markcite{arbvv92}1992).
Unlike clusters, there is no clear correlation between the mass of Fe
in the ICM vs L$_B$ (Arnaud et al 1992). The correlation between
abundance and x-ray luminosity is stronger but we defer that
analysis to Mulchaey et al. (1998) where  we will use the more
accurate $ROSAT$ PSPC fluxes.  

To determine the relative abundance of elements other than Fe  we fit the
x-ray spectra with a Raymond-Smith plasma model with
variable abundances (VRAYMOND). For these fits O, Ne, Mg, Si, and S
(the $\alpha$-elements; A$_\alpha$) vary together. Fe and Ni vary in
unison, and He, C, N, Ca, Ar are fixed at the solar value. 
The ten groups with sufficient signal-to-noise to determine the abundance of
iron and the $\alpha$-elements separately are listed in table 3a. All the
groups in table 3a are well fit with the VRAYMOND model with the exception
of the NGC~5044 group. For NGC~5044 the group gas shows a strong metalicity
gradient and cannot be characterized by a global abundance. It will be
discussed in a later paper. 

Most of the groups listed in table 3a have the abundance of the 
$\alpha$-elements equal to the abundance of iron (within the errors). The 
only two exceptions are the N4104 group and MKW9, which have the iron 
abundance about half that of the $\alpha$-elements. To quantify this, 
the average of the $\alpha$-elements and the average of the iron
abundance $\langle$A$_\alpha\rangle$/$\langle$A$_{Fe}\rangle$ is 1.22,
but if we exclude the ``hot'' groups (kT $>$ 1.5) then the average ratio
drops to 0.96. For the two hot groups (NGC~4104 and MKW~9)
the average $\langle$A$_\alpha\rangle$/$\langle$A$_{Fe}\rangle$ is 2.1.

The spectral resolution of the SIS allows us to fit some of the
individual line complexes. For the groups with greater than
$\sim$10000 counts in the SIS data we allowed O, Si, S, and Fe
to vary independently. The results of the fits are shown in
table 3b. These global fits will include emission from any
hot gas from the central elliptical which might be distinct from the
group gas. The oxygen abundance is not well constrained for any of the groups
examined here, so the large range in the O/Fe ratio (from a high of 6 to a
low of 0.5) may not be particularly meaningful. The S/Fe ratio is
fairly constant ranging from 1.4 -- 0.9 and is, on average, slightly
larger than that seen in clusters $\sim$0.78. 
The Si/Fe ratio in groups varies from 2.7 -- 0.8 with the ratio of Si to
Fe in the hot groups at the high end and the cool groups at the low
end.  Clusters show an average Si/Fe ratio of $\sim$2 
(Mushotzky et al.\markcite{MLA96} 1996). Thus, the average for the groups 
(Si/Fe $\sim$ 1.1) is significantly lower than that of clusters. This trend 
for the  silicon to iron ratio to increase as the temperature (i.e. mass) of
the system increases has previously been noted by Fukazawa et
al. \markcite{FMM96}(1996).

\section{Discussion}

The detection of hot x-ray gas in galaxy groups has opened a window
into the history of the x-ray emitting gas in galaxy systems of mass
$\sim$10$^{13}$-10$^{14}$ M$_{\sun}$. 
The ASCA data allow us to directly probe the elemental abundance of
the hot gas. With this information we can begin to answer some of the
questions about the origin of heavy elements in groups and poor
clusters. Among the issues that need to be resolved are the variation
of abundance with temperature, the origin of the metals, and why 
groups appear to have different enrichment histories than rich clusters.

\subsection{Supernovae in Enrichment of the Diffuse Gas}

The enrichment of the hot gas seen in groups and clusters of galaxies is
widely believed to be from processed stellar material transported
out and mixed with the hot gas in the system. The favored models
contain a large number of type I and type II SNe early in
the lifetime of the cluster, when elliptical galaxies are forming stars
at a rapid rate (Renzini et al. \markcite{RCD93}1993). 
Since type I SNe will produce mostly iron and type II SNe will produce
the $\alpha$-process elements in addition to iron, high resolution
x-ray spectroscopy can, in principle, be used to determine the relative
rate of type I and type II SNe over the enrichment history of the
diffuse hot gas. However, the amount of iron produced in type II SN
events is uncertain (Loewenstein \& Mushotzky \markcite{LM96}1996). 
The ratio of A$_\alpha$ to A$_{Fe}$ measured in clusters, $\sim$2,
favors the majority of the metals originating from type II SNe ejecta
(Mushotzky et al. \markcite{MLA96} 1996; Loewenstein \& Mushotzky
\markcite{LM96} 1996). In our group sample, the ratio of A$_\alpha$ to
A$_{Fe}$ is $\sim$1, implying that type II SNe are less prevalent in
the early evolution of groups than in rich clusters. 

However, interpreting this ratio is further complicated by
the possibility of multiple star formation events in the galaxy. If the early
wind phase of the galaxy is predominately powered by type II SNe and later
star formation episodes are dominated more and more by type I SNe then the
detailed star formation history of the system is needed to interpret the
ratio of A$_\alpha$ to A$_{Fe}$ with confidence. If we take the metal
abundances at face value then
the relative abundance of the $\alpha$-process elements to iron indicates
that for groups both type I $and$ type II SNe contribute significantly to
the enrichment of the diffuse gas. This is in contrast to clusters where
type II SNe seem to dominate the enrichment of the diffuse gas (Mushotzky et
al \markcite{MLA96} 1996, but see Gibson, Loewenstein and Mushotzky
\markcite{GLM97} 1997 for a discussion of the wide range of possibilities).

The ratio of $\alpha$-elements to iron is very sensitive to the 
supernova model used in the calculation. 
The expected abundances relative to iron for various supernova models have
been tabulated by Loewenstein \& Mushotzky \markcite{LM96}(1996)(table 4).
The data for the hot groups show that the Si/Fe ratio is $\sim$2.5 while 
for the cooler groups it is $\sim$1. This, along with the average S/Fe
ratio of $\sim$0.78, indicates that the Woosley \& Weaver \markcite{WW95}(1995) 
models with zero abundances (WW0) and models with zero abundances and
enhanced SNe for M $>$ 25 M$_{\sun}$ (WW0e) fit the observed ratios
best. None of the models they tabulate fit the high O/Fe
ratio seen in the NGC~4104 group, but the oxygen abundance is not 
well-constrained by the ASCA data. The low O/Fe ratio seen in the
cool groups, if confirmed, would be consistent with the zero abundance
models. 

The mass to light ratio of various elements is a powerful tool for
determining the type of SNe events which have enriched the hot gas in
groups and clusters (Renzini et al. \markcite{RCD93}1993).
The mass to light ratio for O, Si, and Fe using various SNe models 
has been tabulated by Loewenstein \& Mushotzky\markcite{LM96}(1996) to
determine the best scenarios for the enrichment of the hot gas in
clusters. Here we examine the mass to light ratios of the groups in our
sample to determine the enrichment history of groups.

The element that is best determined in these x-ray spectra is iron.
The M$_{Fe}$/L$_B$ for the groups ranges 
from (0.5 -- 54.5 )$\times$10$^{-5}$ in solar units (table 4). 
This is much lower than the typical cluster value of 
$\sim$0.02 M$_{\sun}$/L$_{\sun}$. Examining table 3a of Loewenstein \&
Mushotzky \markcite{LM96} 1996, we find that only the models with 
very steep IMF (x=2.35) can fit the observed data. 
But because type I and type II SNe can eject iron into the ISM,
this element is  not well suited for determining the type of early SNe
events that enrich the hot gas.
We have determined the mass to light ratio for individual
$\alpha$-elements in four groups. The oxygen mass to light ratios
range from (2.4 -- 21.8 )$\times$10$^{-3}$M$_{\sun}$/L$_{\sun}$
while the silicon mass to blue light ratio is between 
(3.7 -- 18.2 )$\times$10$^{-4}$M$_{\sun}$/L$_{\sun}$.
The O/L$_B$ and Si/L$_B$ ratios are consistent with IMF's with a slope
steeper than about x = 1.8. But when we include the M$_{Fe}$/L$_B$
ratios only the models with zero initial stellar abundance and 
enhanced SNe rate can reproduce the elemental mass to light ratios in
these groups.

\subsection{Mass Loss from Galaxy Groups}

The $ASCA$ data show that groups have less metals in their hot
gas than do clusters per unit blue light. This low metal mass to light
ratio is due both to a low overall abundance and to a relatively lower
hot gas to total mass ratio. The low ratio seen in the groups could be the
result of several different processes. 
One could be that the metal production per unit light may vary with
environment e.g. with local density.  A second possibility is that
groups lose large amounts of metals during their evolution (White 
\markcite{W91} 1991; David et al.\markcite{DAFJ90} 1990; David et
al.\markcite{DFJ91} 1991;  Renzini \markcite{R97} 1997). 
We examine both of these scenarios below. 

If environment is the key then there must be a feedback process between
the conversion of gas into stars and the large scale environment. Such an
effect is indicated by the morphology density relation (Dressler 1980).
Because the rich clusters are much larger density perturbations than the 
poor groups such a process might be related to a greater efficiency of star
production from gas in overdense perturbations that will become the rich
clusters. However, there is no indication of any difference in the stellar 
populations of ellipticals that reside in clusters or in the poor groups. 
Because the total light of the groups is dominated by spheroidal systems one 
suspects that the metal production rate is also dominated by spheroidals. 
Thus, if local density is the controlling factor then the conversion of
gas into  stars and therefore into metals must somehow have produced "normal"
stellar populations. This local process must also result in a lower
metal to blue light ratio than for similar stellar populations in rich
clusters. 

The proposition that groups lose most of their enriched gas is consistent 
with models (Arnaud et al.\markcite{arbvv92} 1992; Renzini et 
al.\markcite{RCD93} 1993) in which the  
production of metals creates enough energy to produce a large scale wind.
The additional average energy per particle produced by metal formation is
roughly  equivalent to the binding energy of the groups (Loewenstein
and Mushotzky \markcite{LM96} 1996). Thus the ejection of metals from
groups is not unexpected. Additionally, if groups form 
from smaller subunits and if the infall velocity is thermalized this
can provide sufficient energy to  eject the IGM (Gnedin \&
Ostriker\markcite{GO97} 1997). 

Further evidence in support of the wind hypothesis comes from the metal
mass to light ratios. Because clusters and groups
of galaxies are not closed systems and can exchange gas with the
surrounding environment (Ciotti et al. \markcite{CDP91} 1991 and 
the above discussion), the ratio of mass of
metals in the gas to the total blue luminosity of the galaxies
is a more useful quantity for determining the enrichment history
of the gas (Renzini et al.\markcite{RCD93} 1993) than the relative
abundances of the individual elements. 

For a large data base of clusters, Arnaud et al. \markcite{arbvv92}
(1992)  found that in the ICM the M$_{Fe}$ was proportional to the total blue
luminosity, and that the trend was for cooler clusters to have
a higher abundance. However, this trend is rather shallow and
Renzini et al \markcite{RCD93} (1993) adopted a typical value of
M$_{Fe}$/L$_B$ = 0.01 -- 0.02 M$_{\sun}$/L$_{\sun}$.
For groups the ``typical'' value of M$_{Fe}$/L$_B$ is $\sim$0.0001 but
as can be seen in figure 2 the value of M$_{Fe}$/L$_B$ drops rapidly
as the group temperature drops. Therefore, it is likely that the groups must
lose most of the enriched gas that the galaxies in the group can
produce (see the discussion above).
This also indicates that for hierarchical clustering the poor
groups we see today cannot be the building blocks of the richer
systems unless the ejected gas is globally bound to a collection
of merging systems.

The global enrichment of hot group gas varies with temperature in the opposite
sense than is seen in clusters; in clusters the metal abundance increases with
decreasing temperature. For groups with temperatures less than 1.5--2 keV the
trend is increasing metal abundance with increasing temperature. Figure 1 shows
the group data along with a solid line showing the trend for clusters
(Arnaud et al.\markcite{arbvv92} 1992). The break from the abundance predicted
from clusters at $\sim$1.5 keV is consistent with the lower mass
systems  losing part of the metal rich gas ejected from the galaxies
(Mulchaey et al. \markcite{MDM93} 1993; Renzini et al.\markcite{RCD93}
1993; Davis et al.\markcite{DMM96} 1996). If current
models of elliptical evolution are correct then the least massive
(coolest systems) must lose most of their processed material.

The observation that in clusters the M$_{Fe}$/L$_B$ is fairly constant is
an indication that massive clusters are probably closed systems and
have not lost or accreted metals from outside the cluster; or, if they
have done so the process must have a very narrow range of allowed
variance. However, the fact that the average cluster abundance does
show variations (Scharf \& Mushotzky\markcite{SM97} 1997) indicates
that there is true variation from object to object. If clusters were
losing mass via winds (or some such mechanism) then the less massive
(cooler) systems would lose more gas and the M$_{Fe}$/L$_B$ would be
smaller for cooler systems. Similarly, if clusters were accreting
primordial gas from outside the cluster then the clusters accreting
the most gas (the most massive systems) would have their metals diluted
more, resulting in a trend in M$_{Fe}$/L$_B$. Since only a mild trend is
observed in clusters it is assumed that clusters are closed systems.
The simplest high density hierarchical models form clusters at low
redshift from the merger of groups. This is not consistent with our
data since the cluster properties would be the average of the group
properties and our results show that the average group has lower
M$_{Fe}$/L$_B$, lower metal abundance and a lower gas mass fraction
than the rich clusters. However, more complex models in which the
timescale for metal production, the collapse of high density
perturbations, and the possibility of a low density universe are
included, can be made consistent with our results.

For groups, strong downward trends are seen in the M$_{Fe}$/L$_B$ data.
Because groups of galaxies show a strong trend of decreasing abundance
with decreasing temperature (mass) then using the arguments above it
seems that groups of galaxies show evidence for mass loss from winds.
So, unlike clusters, groups are not closed systems.
While mass loss from groups can explain the decreasing trend seen in
the M$_{Fe}$/L$_B$ ratio there could be other explanations.
One possible method of altering the M$_{Fe}$/L$_B$ ratio between systems
is to change the initial mass function. The consistency of M$_{Fe}$/L$_B$
in clusters indicates that the initial mass function (IMF) does not change
much from cluster to cluster (Renzini et al. \markcite{R97} 1997; Wyse
\markcite{W98} 1998).
To adjust the IMF for groups only to explain the M$_{Fe}$/L$_B$ ratio
seems rather contrived given the lack of evidence and we don't
consider this a likely explanation.

\subsection{Evolution of the Group Gas}

The global x-ray properties of groups are consistent with groups being
scaled-down versions of clusters (Price et al. \markcite{P91}1991; Henry et
al. \markcite{H95}1995; Mulchaey \& Zabludoff \markcite{MZ98}
1998). But, the wide variations seen the abundance, abundance  
ratios and M$_{Fe}$/L$_B$ in group gas argues strongly that the
evolution of group gas is not simply a scaled down version of cluster
evolution. 
In clusters the primary source of metals is thought to be from the
early evolution of elliptical galaxies. Proto-elliptical galaxies are
thought to undergo a burst of star formation and subsequently lose
roughly half of their original mass in winds driven by supernova
explosions (David et al. \markcite{DJFD94} 1994). Analysis of ASCA data
indicate that these 
early winds are driven by type II SNe (Loewenstein \& Mushotzky
\markcite{LM96} 1996). Later, as the progenitors of the type II events
are not replenished, type I SNe continue to drive winds, albeit with
a much diminished mass loss rate.

Winds blown from elliptical galaxies can reach speeds of $\sim$1000
km/s and for systems less massive than $\sim$3$\times$10$^{13}$M$_{\sun}$
the wind will not be bound to the system.
Clusters, with masses $\sim$10$^{14}$--10$^{15}$ M$_{\sun}$, can
retain the enriched gas blown off by elliptical galaxies. But,
because group masses show a range of $\sim$10$^{12}$--10$^{13}$ M$_{\sun}$
(Mulchaey et al.\markcite{MDM96} 1996) it is not
surprising to see that groups may retain only some part of that enriched gas.
This can also explain the difference in the
$\langle$A$_\alpha\rangle$/$\langle$A$_{Fe}\rangle$ ratio. If the early
winds from ellipticals are lost by the groups then they also lose the
material enriched primarily by type II SNe. Currently ellipticals
are predicted to have mass loss rates of $\sim$1.5$\times$10$^{-11}$
M$_{\sun}$ yr$^{-1}$ L$_B$ (Faber \& Gallagher\markcite{FG76} 1976). 
This can produce about 10$^6$ M$_{\sun}$ of iron in 10$^{10}$ years 
for a typical group luminosity of 2$\times$10$^{11}$ L$_B$ which could supply
only 10\% of the iron seen in groups today. Any winds from ellipticals are
currently driven by type I SNe, and the abundance ratios in groups
indicate that type I SNe play a larger role in enriching the group gas
than in enriching cluster gas. However, the presence of $\alpha$-elements
shows that groups have retained some of the type II SNe ejecta and
this could provide a source of additional iron and raise the iron mass
to the present day values of between 10$^{7}$--10$^{8}$ M$_{\sun}$.

\section{Baryonic fraction of the Ejected Material and the Intergalactic Medium}

If, as we argue above, most of the enriched intra-group medium has
been ejected then this indicates that groups can pollute the
primordial gas in the intergalactic medium. 
Simulations show that during the early evolution of ellipticals they
lose about half of their total mass as supernova induced winds (David
et al. \markcite{DFJ91} 1991). 
We can estimate the total fraction of baryons ejected from groups
using the data in Mulchaey et al. \markcite{MDM96}(1996) and in Fukugita
et al. \markcite{FHP98}(1998).
If we assume that the galaxy mass has decreased by a factor of two due
to winds and that this extra enriched gas has been ejected from the
group, we find that $\Omega_B\sim$0.008h$^{-1.5}_{50}$ where
$\Omega_B$ is the fraction that the baryons contribute to the critical
density in an Einstein-de Sitter universe. This is less
than the value that Fukugita et al. find for cool plasma in low 
surface density clouds outside of groups, clusters and galaxies. 
Converting the value of $\Omega$ given by Fukugita et
al.\markcite{FHP98} (1998) for the Lyman-$\alpha$ forest and the
damped absorbers to H$_0$ of 50 yields a value of 0.024. So, if groups
do eject gas at early epochs, then they may contribute gas to the
Lyman-$\alpha$ forest and can account for about 1/3 of the gas seen.

\section{Conclusions}

We have analyzed the x-ray data for seventeen groups of galaxies
observed by ASCA. We have determined the global temperature
and abundance for each of the groups. We have also determined the
abundance of $\alpha$-elements for ten of the
groups and investigated the abundance of O, S, Si, and Fe for
several systems.

The global temperature  of the groups are all $\sim$1 keV but the
metal abundance can vary widely from $<$0.05 -- 0.55 solar. We find
that the metal abundance in groups drops rapidly as the temperature drops.
The best fit correlation shows that A$\propto$T(keV)$^{+2.5}$ and is
in the opposite direction than the correlation found for clusters.
The mass of metals per unit blue light also drops rapidly as the group
temperature drops. We interpret this as evidence for winds in the
early history of the group. 

Further evidence for group-scale winds comes from fitting the 
abundances of the $\alpha$-elements and iron separately.
We find that for cool groups (kT $<$ 1.5 keV) the
$\langle$A$_\alpha\rangle$/$\langle$A$_{Fe}\rangle$ ratio is $\sim$1
but for hotter groups it is $\sim$2. This strongly suggests that,
unlike clusters, groups lose most of the gas enriched by type II SNe
during the early life of an elliptical galaxy. This has strong
implications for the existence of a metal enriched intergalactic
medium in the low redshift universe. Simple calculations indicate
that most of the baryons have been ejected from groups and thus 
groups contribute to a putative metal rich intergalactic medium.

The spectral resolution of the ASCA SIS has provided unique insight into the
evolution of groups of galaxies. However, future work will require
better spatial resolution and sufficient spectral resolution to determine
elemental abundances in a sample of groups: a project well suited
to AXAF.

\acknowledgments
We would like to thank our referee for helpful comments
which improved the paper. 
This research  made use of the HEASARC, NED, and SkyView databases.
Partial support for this project was provided by NASA grants NAG
5-3321, NAG 5-3529 and NAG 6-6143. DSD's research at M.I.T. is
supported in part by the AXAF Science Center as part of Smithsonian
Astrophysical Observatory contract SVI--61010 under NASA Marshall
Space Flight Center. 

\newpage
\onecolumn

\newpage
\onecolumn
\begin{table}
\caption{Basic Data for Poor Groups}
\begin{tabular}{lcclll}
\tableline
\tableline
Group&R.A.&Dec.&Velocity&$\sigma$&L$_{\rm b}$\\
      &J2000&J2000&km s$^{-1}$&km s$^{-1}$&10$^{11}$M$_\odot$\\
\tableline
NGC~2300&07 32 20&+85 42 31&\phantom{0}2074&251&1.1\\
NGC~2563&08 20 36&+21 04 00&\phantom{0}4890\tablenotemark{a}&336&3.1\\
HCG~42&09 57 52&$-$19 23 56&\phantom{0}3990&214\tablenotemark{b}&2.7\\
NGC~3258&10 28 54&$-$35 36 22&\phantom{0}2852&176&6.7\\
HCG~51&11 22 21&+24 17 35&\phantom{0}7740&240\tablenotemark{b}&3.7\\
HCG~57&11 35 17&+22 15 28&\phantom{0}9120&269\tablenotemark{b}&5.1\\
NGC~4104&12 06 39&+28 10 18&\phantom{0}8480&546\tablenotemark{c}&4.8\\
HCG~62&12 50 30&$-$08 55 59&\phantom{0}4110&288\tablenotemark{d}&0.9\\
NGC~4325&12 23 07&+10 37 10&\phantom{0}7558&265\tablenotemark{a}&2.4\\
NGC~5044&13 14 23&$-$16 32 04&\phantom{0}2459&360&1.7\\
RGH~80&13 18 06&+33 27 \phantom{00}&11098&467&1.8\\
NGC~5129&13 24 10&+13 58 33&\phantom{0}7096&208&4.5\\
MKW~9&15 32 29&+04 40 54&11253&336\tablenotemark{c}&7.7\\
NGC~6329&17 14 15&+43 41 04&\phantom{0}8223\tablenotemark{e}&\nodata&1.5\\
Pavo&20 18 12&$-$70 53 19&\phantom{0}3815\tablenotemark{f}&169\tablenotemark{f}&4.7\\
HCG~92&22 35 58&+33 57 36&\phantom{0}6450&389&4.2\\
Pegasus&23 20 32&+08 11 27&\phantom{0}4197&780\tablenotemark{d}&15.6\\
\tableline
\end{tabular}
\tablenotetext{a}{Zabludoff \& Mulchaey 1997}
\tablenotetext{b}{Hickson et al. 1992}
\tablenotetext{c}{Beers et al. 1995}
\tablenotetext{d}{Fadda et al. 1996}
\tablenotetext{e}{Velocity from the RC3}
\tablenotetext{f}{Bi-weight estimates}
\end{table}
\begin{table}
\tablenum{2}
\caption{Global X-ray Data for Poor Groups}
\begin{tabular}{lcclccr}
\tableline
\tableline
Group&Temperature&Abundance&N$_{\rm h}$/10$^{20}$&N$_{\rm
h}$/10$^{20}$(Gal)&R$_{extract}$&$\chi^2$/dof\\ 
      &keV&Solar&atoms cm$^{-2}$&atoms cm$^{-2}$&$\arcmin$&\\
\tableline
NGC~2300&0.88$^{+0.04}_{-0.05}$&$<$~0.05&4.25$^a$&4.25&25&37/57\\
NGC~2563&1.39$^{+0.05}_{-0.06}$&0.55$^{+0.24}_{-0.17}$&3.92$^a$&3.92&20&356/248\\
HCG~42&0.90$^{+0.03}_{-0.02}$&0.27$^{+0.11}_{-0.07}$&5.32$^a$&5.32\phantom{0}&8&91/78\\
NGC~3258 &1.85$^{+0.23}_{-0.22}$ &0.28$^{+0.21}_{-0.18}$& 6.48$^a$&6.48&13&324/290 \\
HCG~51&1.24$^{+0.04}_{-0.08}$ & 0.41$^{+0.06}_{-0.10}$ & 7.96$^{+3.10}_{-2.42}$&1.27&\phantom{0}7&454/347\\
HCG~57 & 1.15$^{+0.21}_{-0.08}$ & 0.14$^{+0.26}_{-0.09}$ & 1.82$^a$ &1.82&10&327/274 \\
NGC~4104& 2.16$^{+0.15}_{-0.18}$& 0.41$^{+0.10}_{-0.09}$&0.15$^{+3.47}_{-0.15}$&1.69&12&463/459\\
HCG~62&0.95$^{+0.03}_{-0.03}$&0.21$^{+0.03}_{-0.04}$&6.29$^{+5.11}_{-2.50}$&2.69&23&454/419\\
NGC~4325&1.05$^{+0.04}_{-0.03}$&0.44$^{+0.15}_{-0.10}$&$<$9.40&2.23&10&223/196\\
NGC~5044& 1.00$^{+0.02}_{-0.02}$ & 0.24$^{+0.03}_{-0.03}$  & $<$1.3&5.03&30&225/130 \\
RGH~80&1.04$^{+0.02}_{-0.05}$&0.19$^{+0.04}_{-0.04}$&6.80$^{+8.20}_{-6.50}$&1.03&10&348/319\\
NGC~5129&0.87$^{+0.02}_{-0.05}$&0.14$^{+0.05}_{-0.03}$&$<$15.7&1.77&12&299/271\\
MKW~9 &2.44$^{+0.14}_{-0.11}$& 0.29$^{+0.07}_{-0.06}$& $<$0.39&4.18&11&966/752 \\
NGC~6329&1.32$^{+0.06}_{-0.06}$&0.22$^{+0.09}_{-0.05}$&2.14$^a$&2.14&13&306/285\\
Pavo&0.82$^{+0.04}_{-0.11}$&0.09$^{+0.36}_{-0.06}$&0.06$^{+19.70}_{-\phantom{0}0.06}$&5.20&15&303/269\\
HCG~92$^b$ &0.76$^{+0.05}_{-0.05}$&0.11$^{+0.10}_{-0.05}$&$<$7.42&8.00&\phantom{0}4&157/160 \\
Pegasus&1.05$^{+0.02}_{-0.03}$&0.34$^{+0.05}_{-0.04}$&4.40$^{+5.9}_{-3.35}$&5.03&23&480/474\\
\tableline
\end{tabular}
\tablenotetext{}{$^a$Parameter held constant during the fit}
\tablenotetext{}{$^b$Power law component was included in the fit}
\end{table}
\clearpage

\begin{table}
\tablenum{3a}
\caption{Global Spectral Fits with Variable Abundances}
\begin{tabular}{lcccll}
\tableline
\tableline
Group&Temperature&A$_\alpha$&A$_{\rm Fe}$&N$_{\rm h}$/10$^{20}$&$\chi^2$/dof\\
\tableline
      &keV&Solar&Solar&atoms cm$^{-2}$&\\
\tableline
NGC~4325&1.06$^{+0.02}_{-0.06}$&0.31$^{+0.23}_{-0.13}$&0.42$^{+0.14}_{-0.09}$&$<$24.70&223/196\\
NGC~5129&0.87$^{+0.02}_{-0.05}$&0.20$^{+0.26}_{-0.14}$&0.16$^{+0.06}_{-0.04}$&$<$24.1&297/271\\
NGC~4104& 1.93$^{+0.13}_{-0.19}$&0.80$^{+0.24}_{-0.20}$&0.37$^{+0.11}_{-0.12}$&0.17$^{+4.20}_{-0.17}$&463/459\\
HCG~62&0.95$^{+0.03}_{-0.03}$&0.23$^{+0.08}_{-0.07}$&0.22$^{+0.04}_{-0.04}$&5.96$^{+5.94}_{-2.30}$&454/419\\
NGC~5044\tablenotemark{1}&1.01 &0.30&0.36&6.5& \nodata \\
RGH~80&1.02$^{+0.05}_{-0.05}$&0.28$^{+0.16}_{-0.08}$&0.20$^{+0.05}_{-0.06}$&6.88$^{+9.40}_{-5.63}$&335/317\\
MKW~9&2.18$^{+0.11}_{-0.09}$&0.48$^{+0.15}_{-0.13}$&0.24$^{+0.07}_{-0.05}$&$<$7.47&954/748\\
Pavo&0.77$^{+0.07}_{-0.05}$&0.10$^{+0.20}_{-0.09}$&0.17$^{+0.88}_{-0.04}$&$<$56.26&301/265\\
NGC~6329&1.32$^{+0.04}_{-0.07}$ &0.22$^{+0.20}_{-0.13}$&0.24$^{+0.07}_{-0.05}$&$<$10.7&298/281\\
Pegasus&1.04$^{+0.03}_{-0.02}$&0.35$^{+0.41}_{-0.09}$&0.33$^{+0.17}_{-0.04}$&5.30$^{+7.9}_{-5.3}$&480/474\\
\tableline
\end{tabular}
\tablenotetext{1}{Fit was too poor to obtain errors for the parameters}
\end{table}
\clearpage

\begin{deluxetable}{lcllllllr}
\footnotesize
\tablecolumns{9}
\tablenum{3b}
\tablecaption{Group Abundances for Individual Elements}
\tablehead{
\colhead{Group}&\colhead{Radius}&\colhead{Temperature}&\colhead{A$_{o}$}&
\colhead{A$_{Si}$}&\colhead{A$_{S}$}&\colhead{A$_{Fe}$}&
\colhead{N$_{\rm h}$/10$^{20}$}&\colhead{$\chi^2$/dof}\\
\colhead{}&\colhead{(arcmin)}&\colhead{(keV)}&\colhead{(Solar)}&\colhead{(Solar)}&
\colhead{(Solar)}&\colhead{(Solar)}&\colhead{(atoms cm$^{-2}$)}&\colhead{}\\
}
\startdata
NGC~4104&0-12&1.79$^{+0.30}_{-0.18}$&1.87$^{+3.55}_{-0.91}$&
0.75$^{+0.46}_{-0.21}$&0.43$^{+0.26}_{-0.24}$&0.31$^{+0.21}_{-0.08}$&6.59$^{+6.08}_{-4.67}$&
176/172\nl
\tablevspace{2pt}
\tableline
\tablevspace{2pt}
HCG~62&0-15&1.00$^{+0.02}_{-0.03}$&$<$0.16&
0.32$^{+0.07}_{-0.08}$&0.28$^{+0.17}_{-0.15}$&0.25$\pm$0.04&$<$12.6&443/411\nl
\tablevspace{2pt}
\tableline
\tablevspace{2pt}
MKW~9&0-11&2.23$^{+0.22}_{-0.11}$&$<$1.02&
0.64$^{+0.20}_{-0.16}$&0.33$^{+0.18}_{-0.20}$&0.24$^{+0.09}_{-0.06}$&$<$7.98&932/744\nl
\tablevspace{2pt}
\tableline
\tablevspace{2pt}
Pegasus&0-15&1.06$^{+0.03}_{-0.03}$&0.52$^{+1.07}_{-0.27}$&
0.40$^{+0.24}_{-0.11}$&0.46$^{+0.34}_{-0.23}$&0.34$^{+0.21}_{-0.08}$&$<$11.5&494/472\nl
\enddata
\end{deluxetable}
\clearpage
\begin{figure}
\plotone{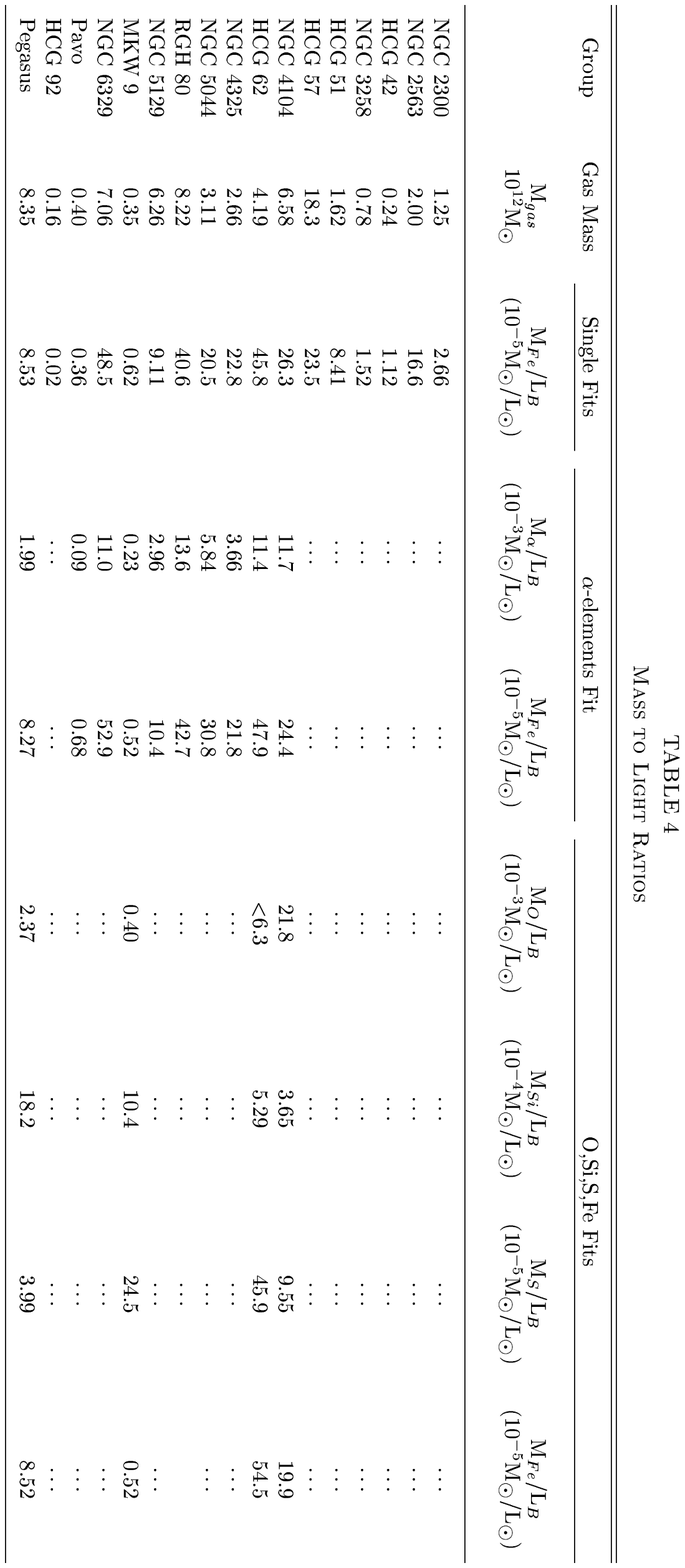}
\end{figure}
\clearpage

\begin{figure}
\figurenum{1}
\plotone{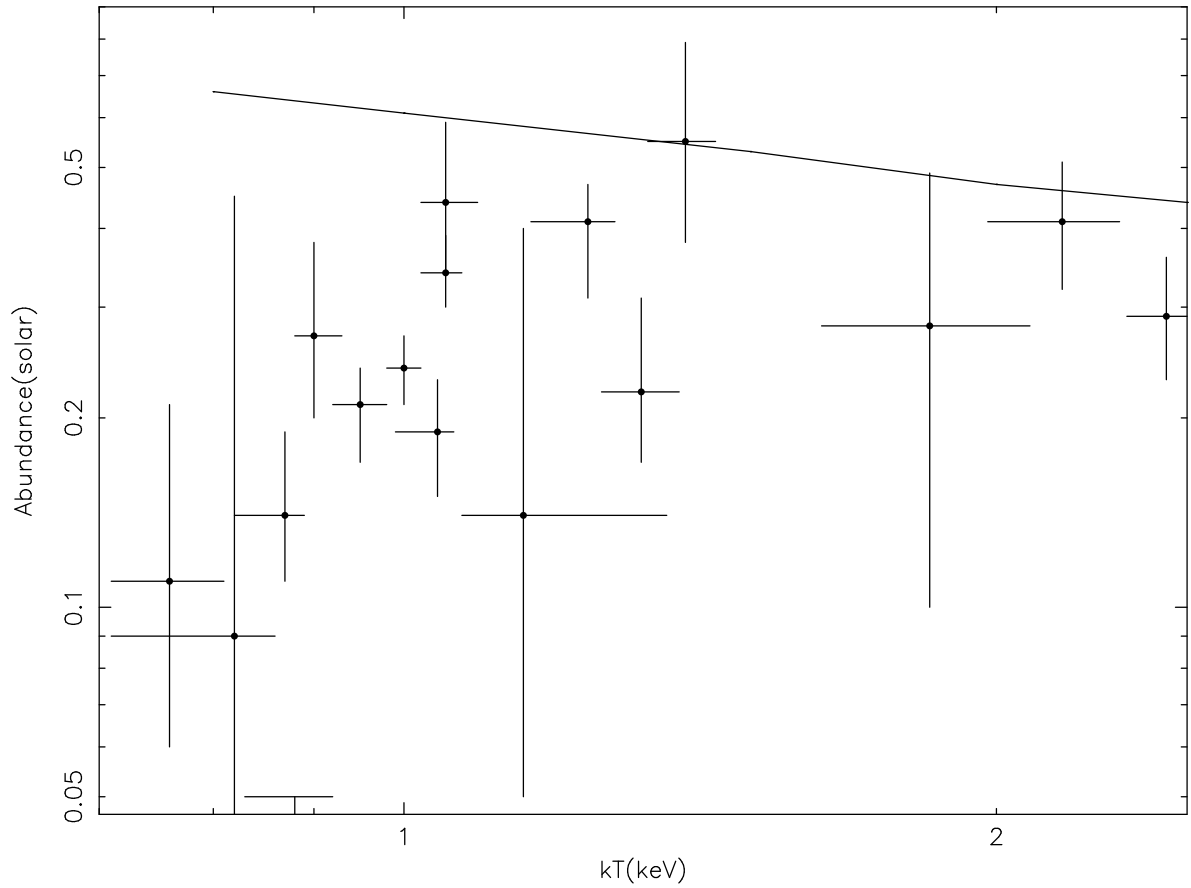}
\caption{The abundance of the hot gas is plotted as a function of the gas
temperature in keV. The error bars are the 90\% confidence range for the
quantities and the solid line shows the best fit for clusters of galaxies from
Arnaud et al. (1992)}
\end{figure}
\begin{figure}
\figurenum{2}
\plotone{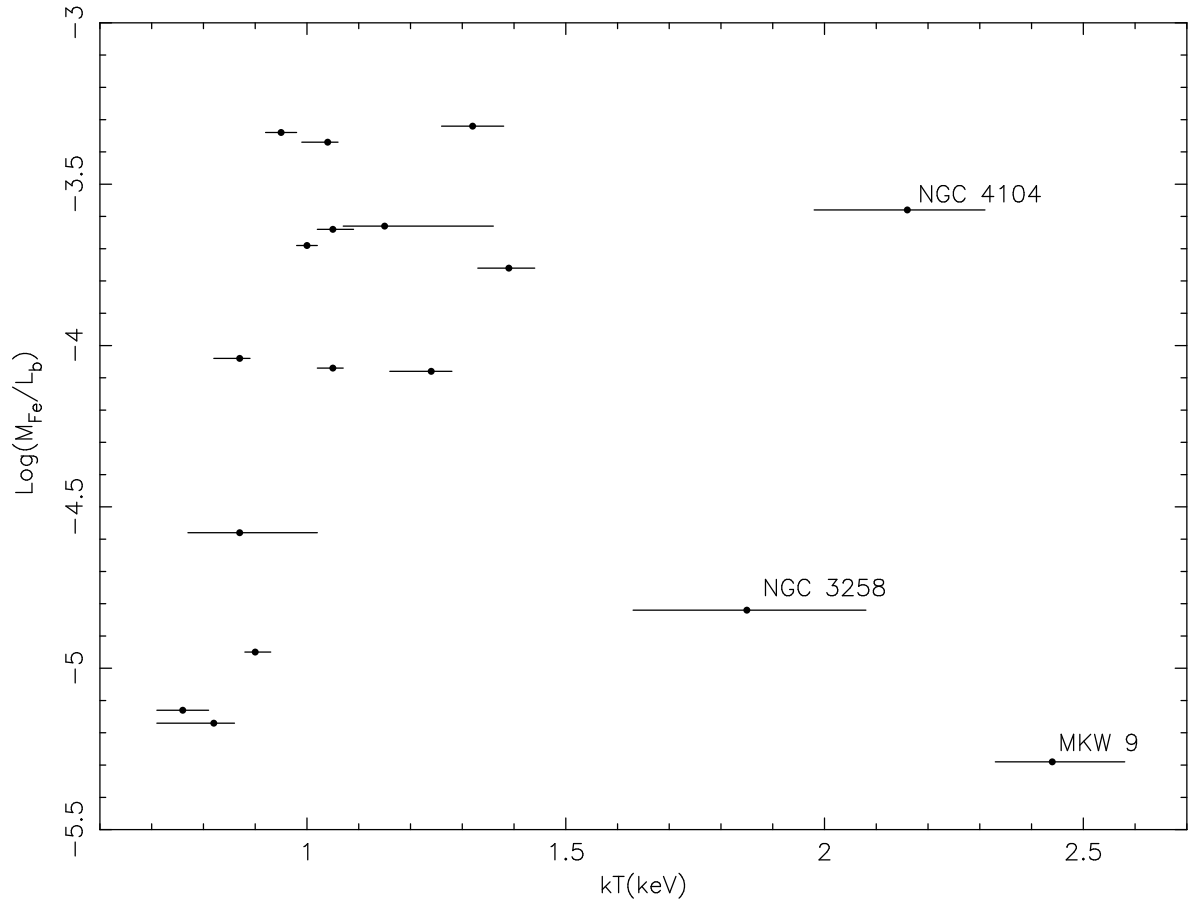}
\caption{The iron mass to light ratio for the groups in this
study is shown as a function of the diffuse gas temperature.}
\end{figure}

\end{document}